\documentclass[a4paper,fleqn,usenatbib]{mnras}

\usepackage[T1]{fontenc}
\usepackage{ae,aecompl}

\usepackage{graphicx}   
\usepackage{amsmath}    
\usepackage{amssymb}    

\newcommand{\be}{\begin{equation}}

\newcommand{\ee}{\end{equation}}
\newcommand{\bea}{\begin{eqnarray}}
\newcommand{\eea}{\end{eqnarray}}
\newcommand{\bef}{\begin{figure}}
\newcommand{\eef}{\end{figure}}
\newcommand{\bce}{\begin{center}}
\newcommand{\ece}{\end{center}}

\def\lsim{\mathrel{\rlap{\lower4pt\hbox{\hskip1pt$\sim$}}
    \raise1pt\hbox{$<$}}}         
\def\gsim{\mathrel{\rlap{\lower4pt\hbox{\hskip1pt$\sim$}}
    \raise1pt\hbox{$>$}}}         
\title[Spin evolution of neutron stars]{Spin evolution of neutron stars in two modes: implication for millisecond pulsars}

\author[Bhattacharyya]{Sudip Bhattacharyya\thanks{E-mail: sudip@tifr.res.in}\\
       Department of Astronomy and Astrophysics, Tata Institute of Fundamental Research, Mumbai 400005, India}
\date{Accepted 2020 December 30. Received 2020 December 29; in original form 2020 November 26}

\pubyear{2015}

\begin{document}
\label{firstpage}
\pagerange{\pageref{firstpage}--\pageref{lastpage}}
\maketitle

\begin{abstract}
An understanding of spin frequency ($\nu$) evolution of neutron stars 
in the low-mass X-ray binary (LMXB) phase is essential to explain the 
observed $\nu$-distribution of millisecond pulsars (MSPs),
and to probe the stellar and binary physics, including the
possibility of continuous gravitational wave emission. 
Here, using numerical computations we conclude that $\nu$
can evolve in two distinctly different modes, 
as $\nu$ may approach a lower spin equilibrium value 
($\nu_{\rm eq,per}$) for persistent accretion
for a long-term average accretion rate ($\dot{M}_{\rm av}$) greater than
a critical limit ($\dot{M}_{\rm av,crit}$), and 
may approach a higher effective spin equilibrium value ($\nu_{\rm eq,eff}$) 
for transient accretion for $\dot{M}_{\rm av} < \dot{M}_{\rm av,crit}$.
For example, when $\dot{M}_{\rm av}$ falls below $\dot{M}_{\rm av,crit}$ 
for an initially persistent source, $\nu$ increases considerably due to 
transient accretion, which is counterintuitive.
We also find that, contrary to what was suggested,
a fast or sudden decrease of $\dot{M}_{\rm av}$ to zero in the last part of the LMXB phase
is not essential for the genesis of spin-powered MSPs, and neutron
stars could spin up in this $\dot{M}_{\rm av}$-decreasing phase.
Our findings imply that the traditional way of $\nu$-evolution computation
is inadequate in most cases, even for initially persistent sources,
and may not even correctly estimate whether $\nu$ increases or decreases.
\end{abstract}
\begin{keywords}
accretion, accretion discs --- methods: numerical --- pulsars: general --- stars: neutron --- stars: rotation --- X-rays: binaries
\end{keywords}



\section{Introduction}\label{Introduction}

A rapidly spinning neutron star, with a measured spin frequency ($\nu$) of $\gsim 100$ Hz, 
is typically known as a millisecond pulsar \citep[MSP; ][]{BhattacharyaHeuvel1991}. 
Such a star is believed to be spun up in its low-mass X-ray binary (LMXB) phase, 
due to the angular momentum transfer by the matter accreted from the companion donor star
\citep{RadhakrishnanSrinivasan1982, Alparetal1982, WijnandsKlis1998, 
ChakrabartyMorgan1998, Archibaldetal2009, Papittoetal2013, Bassaetal2014}.
A subset of neutron stars in the LMXB phase are observed as accretion-powered 
millisecond X-ray pulsars \citep[AMXPs; ][and references therein]{PatrunoWatts2012,DiSalvoSanna2020} and 
nuclear-powered millisecond X-ray pulsars 
\citep[NMXPs; ][and references therein]{Watts2012,Bhattacharyya2020b},
and after this phase some of the stars manifest themselves as spin-powered millisecond
pulsars \citep{BhattacharyaHeuvel1991}.

The neutron star spin evolution can happen by accretion related spin-up and spin-down 
torques and by electromagnetic (EM) and gravitational wave (GW) spin-down torques 
\citep[e.g., ][]{Bildsten1998,BhattacharyyaChakrabarty2017,HaskellPatruno2017,Bhattacharyya2017,Bhattacharyya2020a,Chen2020}.
Hence, the observed $\nu$-value distribution of MSPs can be
very useful to probe the properties and evolution of neutron stars and their binary 
systems, the stellar magnetospheric emission and pulsar wind, which give rise to the
EM torque, and a plausible stellar ellipticty, which should give rise to the GW torque.
Two notable observational aspects of this distribution are it cuts off sharply
$\sim 730$ Hz \citep{Chakrabartyetal2003,Patruno2010,FerrarioWickramasinghe2007}
and MSPs in LMXBs appear to have overall higher $\nu$-values than the post-LMXB phase
spin-powered MSPs \citep{Tauris2012,Papittoetal2014,Patrunoetal2017}. Note that
one needs to understand such observed aspects of the $\nu$-value distribution
in order to use it as a tool to probe the physics of neutron stars and binaries.

The $\nu$-value distribution is primarily caused by the spin evolution 
in the LMXB phase, 
which is generally explained in terms of the spin equilibrium frequency 
($\nu_{\rm eq}$) for persistent accretion, as we briefly describe below.
Note that `persistent accretion' implies that the source
does not show alternate outburst and quiescent cycles with a timescale of
months to years as observed from transients \citep{Linetal2019}, 
but the long-term average accretion rate ($\dot{M}_{\rm av}$) can evolve 
with a timescale of $\sim$ a hundred million years or more.
An accreting neutron star in the LMXB phase spins up and spins down due to the 
accretion through the stellar magnetosphere. The magnetosphere can stop
a thin, Keplerian disc at the magnetospheric radius \cite[e.g., ][]{Wang1996}
\begin{equation}\label{rm}
r_{\rm m} = \xi \left(\frac{\mu^4}{2 G M \dot M^2}\right)^{1/7} ,
\end{equation}
where $\mu$ ($= BR^3$) is the neutron star magnetic dipole moment, $B$ is the
stellar surface dipole magnetic field, $M$ and $R$ are the stellar mass and
radius respectively, $\dot M$ is the instantaneous accretion rate and
$\xi$ is an order of unity constant. Note that, for persistent accretion,
we consider $\dot M = \dot{M}_{\rm av}$.
In the accretion phase, $r_{\rm m}$ is less than the light-cylinder 
radius $r_{\rm lc}$ ($= c/2\pi\nu$) and the corotation radius $r_{\rm co}$, given by
\begin{equation}\label{rco}
        r_{\rm co} = \left(\frac{GM}{4\pi^2 \nu^2}\right)^{1/3},
\end{equation}
and matter and angular momentum are transferred to the neutron star.
On the other hand, in the propeller phase ($r_{\rm co} < r_{\rm m} < r_{\rm lc}$),
the accreted matter is at least partially driven away from the system, and
the star loses angular momentum \citep[see ][]{Watts2012,BhattacharyyaChakrabarty2017}.
In the accretion phase, the star spins up and hence $r_{\rm co}$ decreases,
and in the propeller phase, the star spins down and hence $r_{\rm co}$ increases.
Therefore, $r_{\rm co}$ approaches $r_{\rm m}$, and the neutron star reaches the spin
equilibrium for $r_{\rm co} = r_{\rm m}$, with $\nu$ attaining the
spin equilibrium frequency:
\begin{equation}\label{equilibrium}
\nu_{\rm eq} = \frac{1}{2\pi}\sqrt\frac{GM}{r_{\rm m}^3} =
\frac{1}{2^{11/14}\pi\xi^{3/2}}\left(\frac{G^5 M^5 \dot
        M^3}{\mu^6}\right)^{1/7}.
\end{equation}
After this, $\nu$ typically tracks $\nu_{\rm eq}$.
But if $\dot M$ relatively slowly decreases to zero in the
last part of the LMXB phase, and $\nu$ tracks $\nu_{\rm eq}$ to a very small
value (since, $\nu_{\rm eq} \propto \dot{M}^{3/7}$),
no spin-powered MSP would be created 
\citep[e.g., ][]{Rudermanetal1989,LambYu2005,Tauris2012}.
Therefore, it was indicated that $\dot M$ needs to decrease 
so fast in this last phase of accretion,
that $\nu$ cannot track $\nu_{\rm eq}$ \citep[e.g., ][]{Rudermanetal1989}.
Later, it was reported with an example of numerical computation of 
binary stellar evolution that, 
while $\nu$ decreases when $\dot M$ drastically and rapidly decreases
in the Roche-lobe decoupling phase (RLDP) in the last part of the LMXB phase,
it may not decrease as much as $\nu_{\rm eq}$, 
because $\nu$ may not track $\nu_{\rm eq}$ due to a fast $\dot M$ decrease
\citep{Tauris2012}. It was suggested that this moderate decrease of $\nu$ 
perhaps explains the overall higher $\nu$-values of MSPs in LMXBs relative to 
$\nu$-values of spin-powered MSPs in the post-LMXB phase \citep{Tauris2012}.
In this Letter, we show that $\nu$ could increase
due to transient accretion in the RLDP, conclude that $\nu$ should not attain
a small value even if $\dot M$ relatively slowly decreases to zero at the
end of the LMXB phase, and report complex $\nu$-evolution possibilities and pathways.

\vspace{-0.4cm}
\section{Transient accretion}\label{Transient}

Most neutron star LMXBs, including all AMXPs, accrete matter 
in a transient manner \citep{Liuetal2013,DiSalvoSanna2020}. 
This results in alternate outburst and quiescent phases. The former phase
lasts typically days to weeks when $\dot M$ and the observed X-ray
intensity increase by several orders of magnitude from those in the latter phase, 
which lasts typically months to years \citep[e.g., ][]{YanYu2015}.
These outbursts of accretion are believed to be caused by 
two instabilities:
the thermal instability, for which a small increase in the accretion disc temperature 
$T_{\rm disc}$ causes an additional rise in $T_{\rm disc}$, and the viscous instability,
for which a little increase in $\dot{M}$ leads
to a further increase in $\dot{M}$ 
\citep[see ][ and references therein]{Lasota2001,Doneetal2007}.
The timescale of the former instability is much less than the timescale of the latter.
A thermal instability could occur for $\dot{M}_{\rm av} < \dot{M}_{\rm av,crit}$ 
($\dot{M}_{\rm av,crit}$ is a critical $\dot{M}_{\rm av}$),
when due to the accumulation of matter, the $T_{\rm disc}$ value at a 
certain radius of an initially cold and nonionized disc exceeds
the hydrogen ionization temperature, and hence the opacity increases by a large amount,
photons cannot escape easily anymore, and consequently $T_{\rm disc}$ increases sharply.
This triggers the viscous instability, causing a large increase of $\dot M$,
and hence an outburst happens. But if $\dot{M}_{\rm av}$ is sufficiently
high ($\dot{M}_{\rm av} > \dot{M}_{\rm av,crit}$), so that $T_{\rm disc}$
in the entire disc 
always exceeds the hydrogen ionization temperature, the disc should always be stable,
and accretion should happen persistently.
Therefore, an initially persistent neutron star LMXB can become a transient,
whenever $\dot{M}_{\rm av}$ falls below $\dot{M}_{\rm av,crit}$. Particularly,
this must happen in the last part (e.g., RLDP) of the LMXB phase, 
at the end of which accretion stops.
Note that an expression of $\dot{M}_{\rm av,crit}$, considering the X-ray irradiation of the disc, 
is given by \citep{vanParadijs1996,Kingetal1996,Lasota1997}:
\begin{equation}\label{mdotcrit}
	\dot{M}_{\rm av,crit} \approx 3.2\times10^{15} \left(\frac{M}{M_\odot}\right)^{2/3} \left(\frac{P}{3~{\rm hr}}\right)^{4/3} {\rm ~g~s}^{-1},
\end{equation}
where $P$ is the binary orbital period.

\begin{figure}
\centering
\hspace{-0.64cm}
\includegraphics*[width=9cm,angle=0]{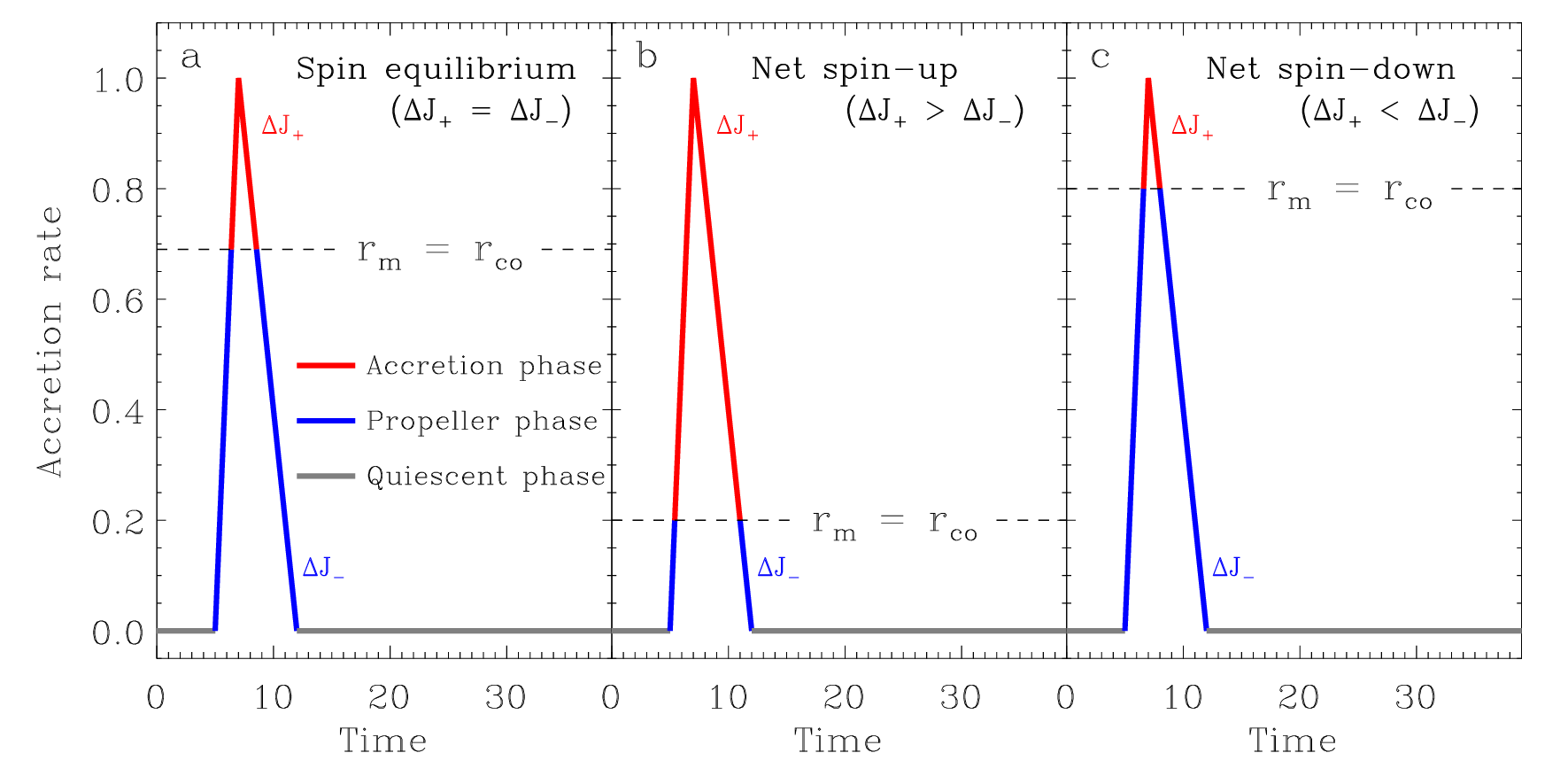}
	\caption{Schematic illustrations of an outburst cycle of a neutron star LMXB \citep[see also ][]{BhattacharyyaChakrabarty2017,Bhattacharyya2017}. The evolution of the instantaneous accretion rate $\dot M$, normalized by the peak accretion rate $\dot{M}_{\rm peak}$, through three phases is shown. Here, we assume a triangular outburst profile, and the time is in an arbitrary unit. During an outburst, the magnetospheric radius $r_{\rm m}$ evolves drastically, while the corotation radius $r_{\rm co}$ remains almost same, and the transition to/from the accretion phase (shown in red) from/to the propeller phase (shown in blue) happens when $r_{\rm m}$ becomes equal to $r_{\rm co}$. Panel (a) is for an overall spin equilibrium of the neutron star, as the total positive angular momentum ($\Delta J_+$) transferred in the accretion phase and the total negative angular momentum ($\Delta J_-$) transferred in the propeller phase are equal. Panels (b) and (c) are for a net spin-up and a net spin-down, respectively (see section~\ref{Transient}).
\label{fig1}}
\end{figure}

A theory and numerical computations of spin evolution for transient sources were reported 
in \citet{BhattacharyyaChakrabarty2017} \citep[see also ][]{Bhattacharyya2017}.
The main difference with persistent sources is, 
as $r_{\rm m} \propto \dot{M}^{-2/7}$ (Equation~\ref{rm}),
during each outburst of weeks to months, $r_{\rm m}$ drastically evolves, while $r_{\rm co}$
($\propto M^{1/3}\nu^{-2/3}$; Equation~\ref{rco}) remains almost the same.
Therefore, the  $r_{\rm co} = r_{\rm m}$ condition is 
almost never satisfied for transient sources, and $\nu$ never tracks 
$\nu_{\rm eq,per}$, which is the spin equilibrium frequency 
(Equation~\ref{equilibrium}) for persistent accretion 
(i.e., $\dot{M} = \dot{M}_{\rm av}$). 
Nevertheless, an approximate or effective spin equilibrium can be achieved for
transient accretion,
if the total angular momentum ($\Delta J_+$) transferred to the neutron star 
in the accretion phase of an outburst is balanced by the total angular momentum
($\Delta J_-$) taken out from the star in the propeller phase of the same outburst
(see Fig.~\ref{fig1}(a)).
The corresponding effective spin equilibrium frequency ($\nu_{\rm eq,eff}$) was
estimated to be 
\begin{equation}\label{equilibriumtran}
	\nu_{\rm eq,eff} = k \nu_{\rm eq,peak} 
	= \frac{k}{2^{11/14}\pi\xi^{3/2}}\left(\frac{G^5 M^5 \dot{M}_{\rm peak}^3}{\mu^6}\right)^{1/7},
\end{equation}
where $k \approx 0.85$ for a triangular outburst profile
\citep[Fig.~\ref{fig1}; ][]{BhattacharyyaChakrabarty2017,Bhattacharyya2017}.
Here, $\nu_{\rm eq,peak}$ is $\nu_{\rm eq}$ (Equation~\ref{equilibrium}) corresponding
to $\dot{M}_{\rm peak}$, which is the $\dot{M}$ of the peak of the outburst.
When $\nu < \nu_{\rm eq,eff}$, a relatively high $r_{\rm co}$ implies a 
sufficiently low $\dot{M}$ corresponding to $r_{\rm m} = r_{\rm co}$, which causes
$\Delta J_+ > \Delta J_-$, and hence the star spins up (see Fig.~\ref{fig1}(b)).
Similarly, for $\nu > \nu_{\rm eq,eff}$, the star spins down (see Fig.~\ref{fig1}(c)).
Therefore, while $\nu$ approaches $\nu_{\rm eq,per}$ for a persistent accretion,
$\nu$ approaches $\nu_{\rm eq,eff}$ for a transient accretion.
Note that, since a typical $\dot{M}_{\rm peak}/\dot{M}_{\rm av}$ value could be
$\sim 10-100$ for transient sources \citep{Burderietal1999}, transient accretion 
should spin up neutron stars to rates several times ($\sim 2-6$ for triangular 
outburst profiles; from Equations~\ref{equilibrium} and \ref{equilibriumtran}) 
higher than can persistent accretion \citep{BhattacharyyaChakrabarty2017}.

In this Letter, we perform spin evolution computations for both persistent
accretion and transient accretion using an evolving $\dot{M}_{\rm av}$.
For computations involving transient accretion, we follow
the method of \citet{BhattacharyyaChakrabarty2017,Bhattacharyya2017},
which was not used for an evolving $\dot{M}_{\rm av}$ earlier.

\vspace{-0.4cm}
\section{Effects of evolving long-term average accretion rate}\label{average}

\subsection{Method}\label{Method}

For our numerical computation of spin evolution, both due to persistent accretion and
transient accretion, we follow the same method described in 
\citet{BhattacharyyaChakrabarty2017}, evolve the star for two billion years,
and use reasonable parameter values for
the purpose of demonstration. For example, we assume the initial values of
$\nu$ and $M$ as 1 Hz and 1.35 $M_\odot$, respectively 
\citep[see ][]{BhattacharyyaChakrabarty2017}, 
and a fixed $B$ value of $10^8$ G. A fixed $B$-value may not be unreasonable 
\citep[see ][]{BhattacharyyaChakrabarty2017}, and is useful to cleanly demonstrate 
the effect of an evolving $\dot{M}_{\rm av}$, which is the aim of this Letter.
Note that the effect of a slightly decaying $B$ in the course of evolution would
be to increase $\nu$ ($\propto B^{-6/7}$ in the equilibrium; 
Equations~\ref{equilibrium}, \ref{equilibriumtran}), 
which does not change our conclusion.

For the spin evolution, we use following expressions of torques 
due to disc--magnetosphere interaction \citep{Rappaportetal2004,BhattacharyyaChakrabarty2017}:
\begin{equation}\label{Torque7}
	N_{\rm acc} = \dot M \sqrt{GMr_{\rm m}} + \frac{\mu^2}{9 r_{\rm m}^3}\left[2\left(\frac{r_{\rm m}}{r_{\rm co}}\right)^3-6\left(\frac{r_{\rm m}}{r_{\rm co}}\right)^{3/2}+3\right]
\end{equation}
for the accretion phase, and
\begin{equation}\label{Torque8}
	N_{\rm prop} = -\eta \dot M \sqrt{GMr_{\rm m}} -\frac{\mu^2}{9 r_{\rm m}^3}\left[3-2\left(\frac{r_{\rm co}}{r_{\rm m}}\right)^{3/2}\right]
\end{equation}
for the propeller phase. $\eta$ is an order of unity positive constant
(we use $\eta = 1$).
In Equations~\ref{Torque7} and \ref{Torque8}, the first term is the accreting 
material contribution, and the second term is the contribution from 
disc--magnetosphere interaction.

For transient accretion, we evolve a neutron star through a series of outburst and 
quiescent phases (see Fig.~\ref{fig1}). 
Note that the  outburst duty cycle (fractional duration) is
$2\dot{M}_{\rm av}/\dot{M}_{\rm peak}$ for triangular outburst profiles
\citep{BhattacharyyaChakrabarty2017}.
While in reality, the 
$\dot{M}_{\rm peak}$-values can have a distribution, and the outburst profiles
can have various irregular shapes, our simple triangular outburst profiles and a fixed 
$\dot{M}_{\rm peak}$-value are useful to cleanly demonstrate the effect of an 
evolving $\dot{M}_{\rm av}$. Note that \cite{Bhattacharyya2017} showed how to
compute a spin evolution for a known $\dot{M}_{\rm peak}$ distribution,
but such a detailed computation would not change our general conclusion.
Besides, a different outburst profile would only imply 
a different $k$-value in Equation~\ref{equilibriumtran}
\citep[e.g., $0.71 \lsim k \lsim 0.85$ for a linear rise and exponential decay, and 
$1 > k \gsim 0.85$ for a flat top; ][]{BhattacharyyaChakrabarty2017},
but would not change our conclusion.

\begin{figure}
\centering
\hspace{-1.0cm}
\includegraphics*[width=8cm,angle=0]{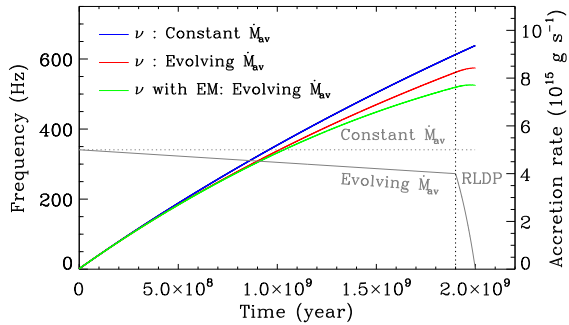}
	\caption{Numerically computed neutron star spin frequency ($\nu$) evolution curves due to transient accretion (see sections~\ref{Method} and \ref{Results1}). The dotted grey horizontal line shows a constant long-term average accretion rate ($\dot{M}_{\rm av}$) of $5\times10^{15}$~g s$^{-1}$. The solid grey curve shows an evolving $\dot{M}_{\rm av}$ with a fast decay (see section~\ref{Method}) in the Roche-lobe decoupling phase (RLDP) on the right of the dotted vertical line. Various $\nu$ evolution curves are for $\dot{M}_{\rm peak} = 5\times10^{17}$~g s$^{-1}$, and based on if electromagnetic (EM) spin-down and $\dot{M}_{\rm av}$ evolution are included not, as mentioned on the plot. This figure shows that $\nu$ can increase in the RLDP (see section~\ref{Results1}).
\label{fig2}}
\end{figure}

Since our aim is to report general results, we do not use $\dot{M}_{\rm av}$
evolution profiles for specific LMXB systems. Rather, we use a parametric formula
(see Fig.~\ref{fig2} for a profile), which is useful and adequate for our purpose. 
One of our aims is to find how $\nu$ evolves in the last part of 
the LMXB phase, when $\dot{M}_{\rm av}$ decreases to zero. This phase could
significantly contribute to the creation of spin-powered MSPs and the $\nu$-distribution
(see section~\ref{Introduction}). Following \cite{Tauris2012}, we 
consider a rapid $\dot{M}_{\rm av}$ decay in this phase,
referred as the RLDP (see section~\ref{Introduction}).
For this fast $\dot{M}_{\rm av}$ decay during the
last $\sim 5$\% time of the LMXB phase \citep[e.g., see ][]{Tauris2012},
we assume the parametric formula: 
$\dot{M}_{\rm av} = a[2-\exp(t/c)]$, where $t$ is time, and $a$ and $c$ are
positive constants.
Besides, to study the effect of the relative values of $\dot{M}_{\rm av}$ and
$\dot{M}_{\rm av,crit}$, we use a slow linear decay of $\dot{M}_{\rm av}$
before the RLDP, and a constant $\dot{M}_{\rm av,crit}$. Note that, while both
$\dot{M}_{\rm av}$ and $\dot{M}_{\rm av,crit}$ could evolve throughout the LMXB phase
in a more complex manner \citep[e.g., ][]{BhattacharyaHeuvel1991,ChenPodsiadlowski2016}, 
our simple assumption is ideal to cleanly demonstrate the main effects of 
$\dot{M}_{\rm av}$ evolution. For the purpose of demonstration, we also
use $\dot{M}_{\rm av,crit} = 10^{16} {\rm ~g~s}^{-1}$, which is reasonable
considering Equation~\ref{mdotcrit} and measured $P$ values of $\sim 1-10$ hr
for AMXPs \citep{DiSalvoSanna2020}.

\vspace{-0.4cm}
\subsection{Spin evolution of transient sources}\label{Results1}

First, we study how $\nu$ evolves if the neutron star accretes transiently
throughout the LMXB phase (see Fig.~\ref{fig2}). For an evolution duration
of two billion years and for the examples given in this figure, $\nu$ remains 
much lower than $\nu_{\rm eq,eff}$ (not shown in the figure) 
for $\dot{M}_{\rm peak} = 5\times10^{17}$~g s$^{-1}$.
Note that an 
$\dot{M}_{\rm av} \sim 5\times10^{15}$~g s$^{-1} - 10^{16}$~g s$^{-1}$
is common for many neutron star LMXBs \citep{LambYu2005}, 
although $\dot{M}_{\rm av}$ could be 
significantly higher in the initial stage of the LMXB phase
\citep[e.g., see ][]{Chenetal2020,Tauris2018}.
The aim here is to find out how $\nu$ evolves
in the RLDP. Fig.~\ref{fig2} shows that $\nu$ can increase in the RLDP, even 
if $\dot{M}_{\rm av}$ decreases drastically, and regardless of the inclusion 
of an additional spin-down (e.g., due to the EM torque 
\citep[see ][]{BhattacharyyaChakrabarty2017} in the quiescent phase; 
see section~\ref{Introduction}). This is not unexpected, as $\nu$ neither approaches
nor tracks $\nu_{\rm eq,per}$
during a transient accretion (see section~\ref{Transient}). However, we note that 
$\nu$ either approaches or tracks $\nu_{\rm eq,eff}$ for such an accretion.
Hence, $\nu$ either increases or decreases due to transient accretion,
both in the RLDP and before the RLDP, depending on whether $\dot{M}_{\rm peak}$
increases or decreases (see section~\ref{Transient} and Equation~\ref{equilibriumtran}).
Therefore, $\nu$ can decrease in the RLDP, if $\dot{M}_{\rm peak}$ decreases.
But, since $\dot{M}_{\rm peak}$ is expected to depend on the binary orbital period $P$
\citep[with an estimated relation of $\dot{M}_{\rm peak} \propto P^{1.79}$; ][]{Lasota2001},
and the change of $P$ is moderate \citep[e.g., ][]{BhattacharyaHeuvel1991},
$\nu$ should not considerably decrease in the RLDP.

\begin{figure}
\centering
\hspace{-1.0cm}
\includegraphics*[width=9cm,angle=0]{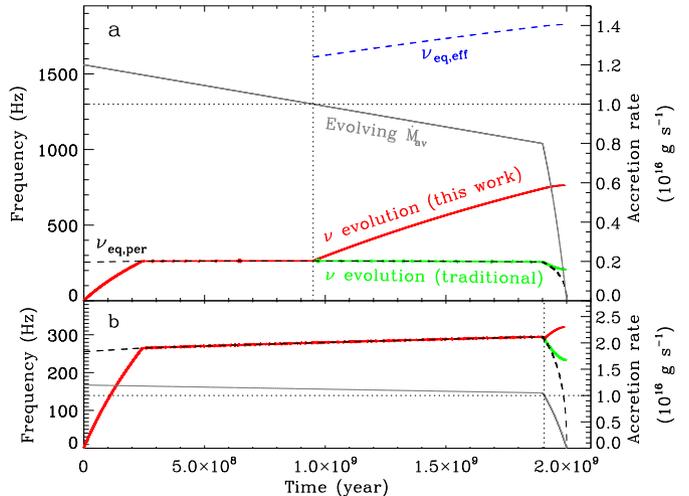}
	\caption{Numerically computed neutron star spin frequency ($\nu$) evolution curves, where the neutron star initially accretes persistently, and then, as the long-term average accretion rate ($\dot{M}_{\rm av}$) decreases below a critical value $\dot{M}_{\rm av,crit}$ (here assumed to be $10^{16}$~g s$^{-1}$, shown by dotted horizontal lines), accretes transiently (see parameter values, the method and a discussion in sections~\ref{Method} and \ref{Results2}). 
	Panel (a): The $\dot{M}_{\rm av}$ curve (solid grey) falls below $\dot{M}_{\rm av,crit}$ before the RLDP. The dotted vertical line marks the time of $\dot{M}_{\rm av} = \dot{M}_{\rm av,crit}$. The solid red curve shows the expected $\nu$-evolution, initially for persistent accretion ($\dot{M}_{\rm av} > \dot{M}_{\rm av,crit}$), when $\nu$ first approaches and then tracks the lower spin equilibrium frequency $\nu_{\rm eq,per}$ (black dashed curve; see section~\ref{Transient}), and then for transient accretion ($\dot{M}_{\rm av} < \dot{M}_{\rm av,crit}$), when $\nu$ approaches the higher effective spin equilibrium frequency $\nu_{\rm eq,eff}$ (blue dashed curve; see Equation~\ref{equilibriumtran}).
Note that, for a sufficiently long transient accretion phase, $\nu$ could evolve to values much higher than those observed (depending on source parameter values), unless there is an additional spin-down torque due to the gravitational wave emisson \citep{BhattacharyyaChakrabarty2017}. We do not include such a torque in this Letter.
The solid green curve shows the traditionally computed $\nu$-evolution for $\dot{M}_{\rm av} < \dot{M}_{\rm av,crit}$, where $\nu$ first tracks $\nu_{\rm eq,per}$, and then does not track it but decreases if $\dot{M}_{\rm av}$ rapidly falls. This panel shows the two distinctly different {\it modes} of $\nu$-evolution (red curve): (i) when $\nu$ approaches and tracks $\nu_{\rm eq,per}$, and (ii) when $\nu$ approaches $\nu_{\rm eq,eff}$.
	Panel (b): Similar to panel (a), but $\dot{M}_{\rm av}$ falls below $\dot{M}_{\rm av,crit}$ in the RLDP, and the $\nu_{\rm eq,eff}$ curve is not shown. This panel shows that a persistent source should become a transient source at least in the RLDP, and then its $\nu$-value increases (see section~\ref{Results2}).
\label{fig3}}
\end{figure}

\vspace{-0.4cm}
\subsection{Spin evolution of persistent sources}\label{Results2}

How does $\nu$ evolve for a persistent source? The neutron star first
spins up relatively quickly towards the spin equilibrium frequency ($\nu_{\rm eq,per}$),
and then tracks it (see sections~\ref{Introduction} and \ref{Transient}, 
and Fig.~\ref{fig3}).
In the RLDP, if $\dot{M}_{\rm av}$ decreases rapidly, $\nu$ decreases
for a persistent source, but cannot track $\nu_{\rm eq,per}$ anymore,
which was shown by \citet{Tauris2012}, and is confirmed in Fig.~\ref{fig3}.
But in reality, a persistent source becomes a transient for
$\dot{M}_{\rm av} < \dot{M}_{\rm av,crit}$, and consequently, $\nu$ increases,
as it does not track the lower spin equilibrium frequency $\nu_{\rm eq,per}$ anymore,
and now approches a higher effective spin equilibrium value $\nu_{\rm eq,eff}$
(see Fig.~\ref{fig3}, and also section~\ref{Transient}). 
Therefore, since even an initially 
persistent source should become a transient at least in the RLDP due to the decrease
of $\dot{M}_{\rm av}$, $\nu$ should increase in this phase (see Fig.~\ref{fig3}(b)).
The condition $\dot{M}_{\rm av} < \dot{M}_{\rm av,crit}$ can also be satisfied 
before the RLDP, and then $\nu$ can drastically increase in the pre-RLDP
and continue to increase in the RLDP (see Fig.~\ref{fig3}(a)). 
These are remarkably different from the currently believed 
spin evolution scenario, and to the best of our knowledge, such a possibility of 
$\nu$-evolution in two distinctly different {\it modes} (see Fig.~\ref{fig3}) 
was not previously demonstrated. 
Moreover, while we do not show a $\dot{M}_{\rm av} < \dot{M}_{\rm av,crit}$ 
to $\dot{M}_{\rm av} > \dot{M}_{\rm av,crit}$
transition in Fig.~\ref{fig3}, such a transition would usually cause a spin-down,
as $\nu$ would approch a lower spin equilibrium value ($\nu_{\rm eq,per}$).

\vspace{-0.4cm}
\section{Discussion and conclusions}\label{Conclusion}

In this Letter, we have demonstrated, for the first time to the best of our knowledge,
that the neutron star spin frequency $\nu$ can approach one of the two 
spin equlibrium frequencies: a lower one ($\nu_{\rm eq,per}$) for persistent accretion
and a higher one ($\nu_{\rm eq,eff}$) for transient accretion. The former
is expected to happen at a higher long-term average accretion rate ($\dot{M}_{\rm av}$)
relative to a critical value ($\dot{M}_{\rm av,crit}$), and the latter 
can happen at a lower $\dot{M}_{\rm av}$ relative to $\dot{M}_{\rm av,crit}$.
So there are two {\it modes} of spin evolution, and the mode leading to the higher
$\nu$ occurs typically for lower $\dot{M}_{\rm av}$ values, which is somewhat
counterintuitive. While we have shown this with examples of a sudden spin-up
with a higher rate for a simple transition from
$\dot{M}_{\rm av} > \dot{M}_{\rm av,crit}$ to $\dot{M}_{\rm av} < \dot{M}_{\rm av,crit}$
(see Fig.~\ref{fig3}),
an opposite transition may lead to a sudden spin-down.
This gives an idea of how a complex $\nu$-evolution curve,
in which $\nu$ evolves by two alternate modes,
may be caused by $\dot{M}_{\rm av}$ and $\dot{M}_{\rm av,crit}$ 
crossing each other multiple times in the LMXB phase.
Even within the transient accretion regime, $\nu$-evolution may
be complex, if $\dot{M}_{\rm peak}$ evolves (section~\ref{Results1}).
This is because $\nu_{\rm eq,eff}$ evolves 
with $\dot{M}_{\rm peak}$ (Equation~\ref{equilibriumtran}), and hence $\nu$
approaches an equilibrium frequency which may change continuously in a complex way.
For example, depending on the evolution of $\dot{M}_{\rm peak}$, $\nu_{\rm eq,eff}$
could evolve to values higher or lower than the $\nu$-value, and hence $\nu$ could increase
or decrease, respectively (see section~\ref{Transient} and Fig.~\ref{fig1}).

We have also particularly studied the spin evolution in the RLDP.
As mentioned above, in the transient accretion regime, 
$\nu$ can spin up or spin down depending on the $\dot{M}_{\rm peak}$-evolution
both in the RLDP and before the RLDP. However, even if $\nu$ decreases
in the RLDP, it should not decrease considerably (see section~\ref{Results1}),
and such a spin-down is not due to the previously
suggested (see section~\ref{Introduction}) tracking of $\nu_{\rm eq,per}$ by $\nu$.
On the other hand, an initially persistent source becomes a transient at least
in the RLDP, as $\dot{M}_{\rm av}$ sufficiently decreases,
and hence $\nu$ should increase, as it now approches a higher effective spin 
equilibrium frequency $\nu_{\rm eq,eff}$ (see section~\ref{Results2}).
Such a $\nu$-evolution (see Fig.~\ref{fig3}) implies that 
a breaking of $\nu$ from $\nu_{\rm eq,per}$ and its subsequent decrease in the RLDP 
\citep[suggested, for example, by ][]{Tauris2012} 
may not be relevant in many cases to determine the final $\nu$-value of the 
LMXB phase. In fact, $\nu$ does not track or approach
$\nu_{\rm eq,per}$ anyway for transient accretion (see section~\ref{Transient}).
Therefore, as the $\dot{M}_{\rm av}$-evolution does not determine the 
$\nu$-evolution, $\nu$ should not attain a low value (e.g., below $\sim 100$ Hz), 
even if $\dot{M}_{\rm av}$ decreases relatively slowly to zero.
This implies that, contrary to what was suggested earlier (see section~\ref{Introduction}),
a fast decrease of $\dot{M}_{\rm av}$ in the last part of the LMXB
phase is not essential for the creation of spin-powered MSPs.

Finally, we note that our findings open up possibilities of many pathways 
by which $\nu$ can evolve, and the traditional $\nu$-evolution computation 
\citep[e.g., ][]{Tauris2012} could work only in special cases, for example, 
when the source remains persistent till the RLDP, and then $\dot{M}_{\rm av}$
decreases very rapidly for $\dot{M}_{\rm av} < \dot{M}_{\rm av,crit}$ in the RLDP.
The evolution of neutron stars via two modes due to persistent 
accretion and transient accretion should significantly affect $\nu$ and other
observed parameter values of MSPs. Such an evolution could be studied in more
detail in the near future using a considerably larger MSP sample, 
acquired, for example, by observations with the Square Kilometre Array 
\citep{Keaneetal2015}.


\vspace{-0.5cm}
\section{Data Availability}

Any other relevant data will be available on request.

\vspace{-0.5cm}


\bibliography{ms}

\bibliographystyle{mnras}

\bsp    
\label{lastpage}
\end{document}